\documentclass{PoS}

\title{Searching for Hybrid Mesons with GlueX}

\ShortTitle{Searching for Hybrid Mesons with GlueX}

\author{\speaker{Sean Dobbs}%
      for the GlueX Collaboration \\
        Florida State University\\
        E-mail: \email{sdobbs@fsu.edu}}

\abstract{Hybrid mesons consist of a quark-antiquark pair bound together by a gluonic field that is in an excited state.  Measuring the spectrum of these states will provide valuable information on the gluonic degrees of freedom of QCD in the quark-confinement regime.
 A rich spectrum of hybrid meson states has been predicted, but only a few experiments have reported evidence of their existence.  The GlueX experiment at Jefferson Lab is designed to search for and measure the spectrum of light-mass hybrid mesons, and it has began its physics run in Spring 2017. For the experiment, a 12 GeV electron beam incident on a diamond radiator is used to produce a linearly-polarized, coherent bremsstrahlung tagged-photon beam with a coherent peak at 9 GeV.  The linearly-polarized photon beam is incident on a proton target located within the hermetic GlueX detector, which can detect many different  final states to which the hybrid mesons are predicted to decay. Measurements with these initial data are discussed, including beam asymmetry measurements, the search for photoproduced $\Xi$ baryons, and near-threshold charm production.}

\FullConference{XVII International Conference on Hadron Spectroscopy and Structure\\
                 25-29 September, 2017\\
                 University of Salamanca, Salamanca, Spain}

\begin{document}

\section{Introduction}

Among the exotic hadrons, those which do not fit into the molds of the $q\bar{q}$ mesons and $qqq$ baryons, the ``hybrid'' mesons are expected to yield particular insight into the nature of quark confinement.  Hybrid mesons are those in which the confining gluonic field directly contributes to the quantum numbers ($J^{PC}$) of the hadron~\cite{hybrids}.  Recent lattice QCD calculations indicate that several hybrid mesons have ``exotic'' quantum numbers, which are not accessible by a quark-antiquark pair, and that have masses accessible at GlueX~\cite{lattice}.  The primary goal of the GlueX experiment is to search for and study these hybrid mesons.  The GlueX experiment is located at the newly-built Hall D at Jefferson Lab, where 12 GeV electrons from the  Continuous Electron Beam Accelerator Facility (CEBAF) are used to generate linearly polarized photons aimed at a hydrogen target surrounded by a nearly hermetic detector. 

Commissioning of the GlueX detector was completed in Spring of 2016, and $\sim80$ hours of physics-quality data were taken.  The first GlueX physics run began in Spring 2017, yielding a factor $\sim8$ increase in the number of events, and consisting of roughly $20\%$ of the first phase of GlueX data taking.  This paper will describe several results using these initial data, including the first studies of meson photoproduction mechanisms at GlueX, which give an initial step towards the full light meson spectroscopy program, and two notable "opportunistic" measurements, studies of $\Xi$ baryons and $J/\psi$ photoproduction.

\section{The GlueX Detector and Beamline}

\begin{figure}[!tb]
\begin{center}
\raisebox{10pt}{\includegraphics[width=3.2in]{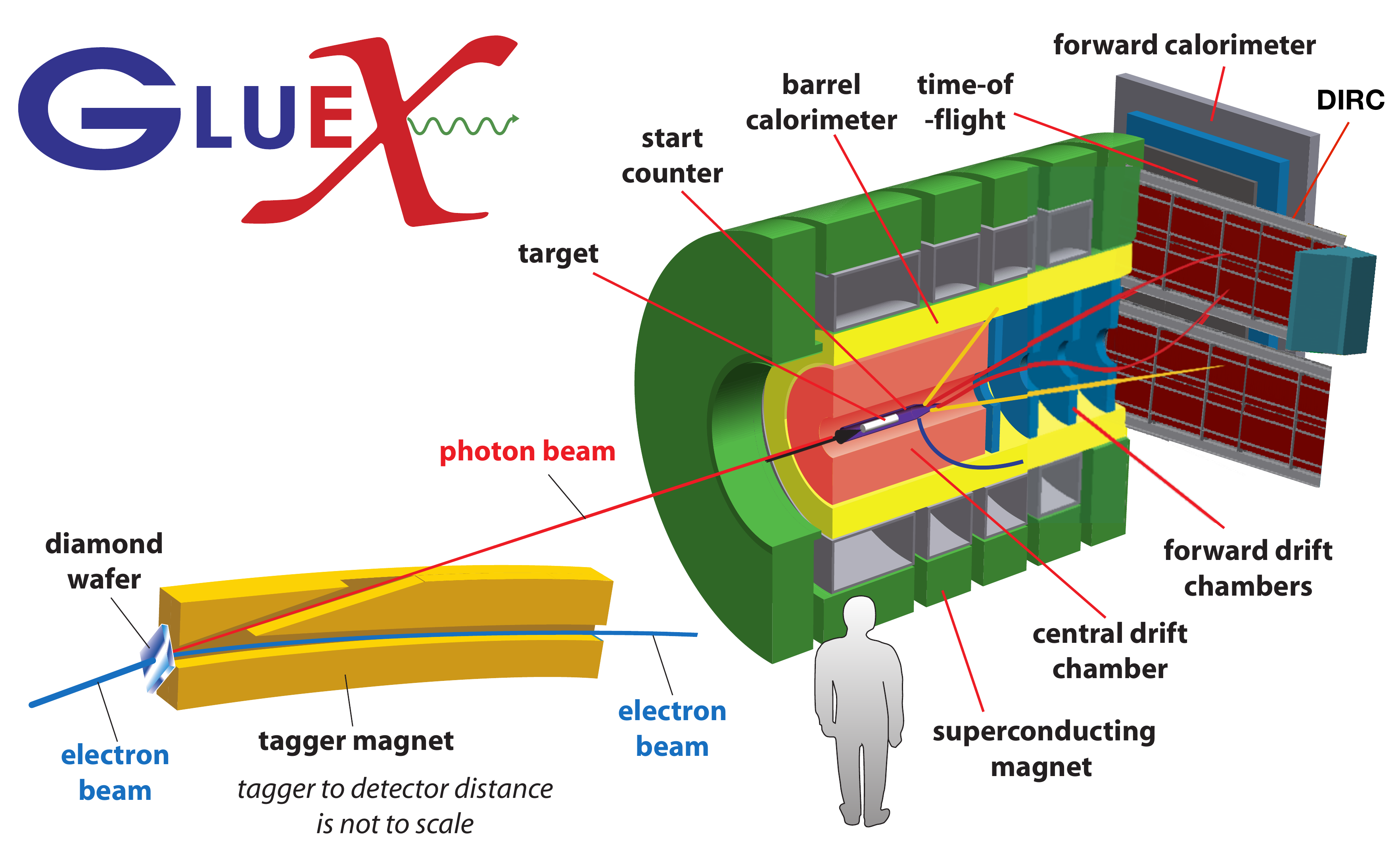}}
\hspace{10pt}
\includegraphics[width=2.5in]{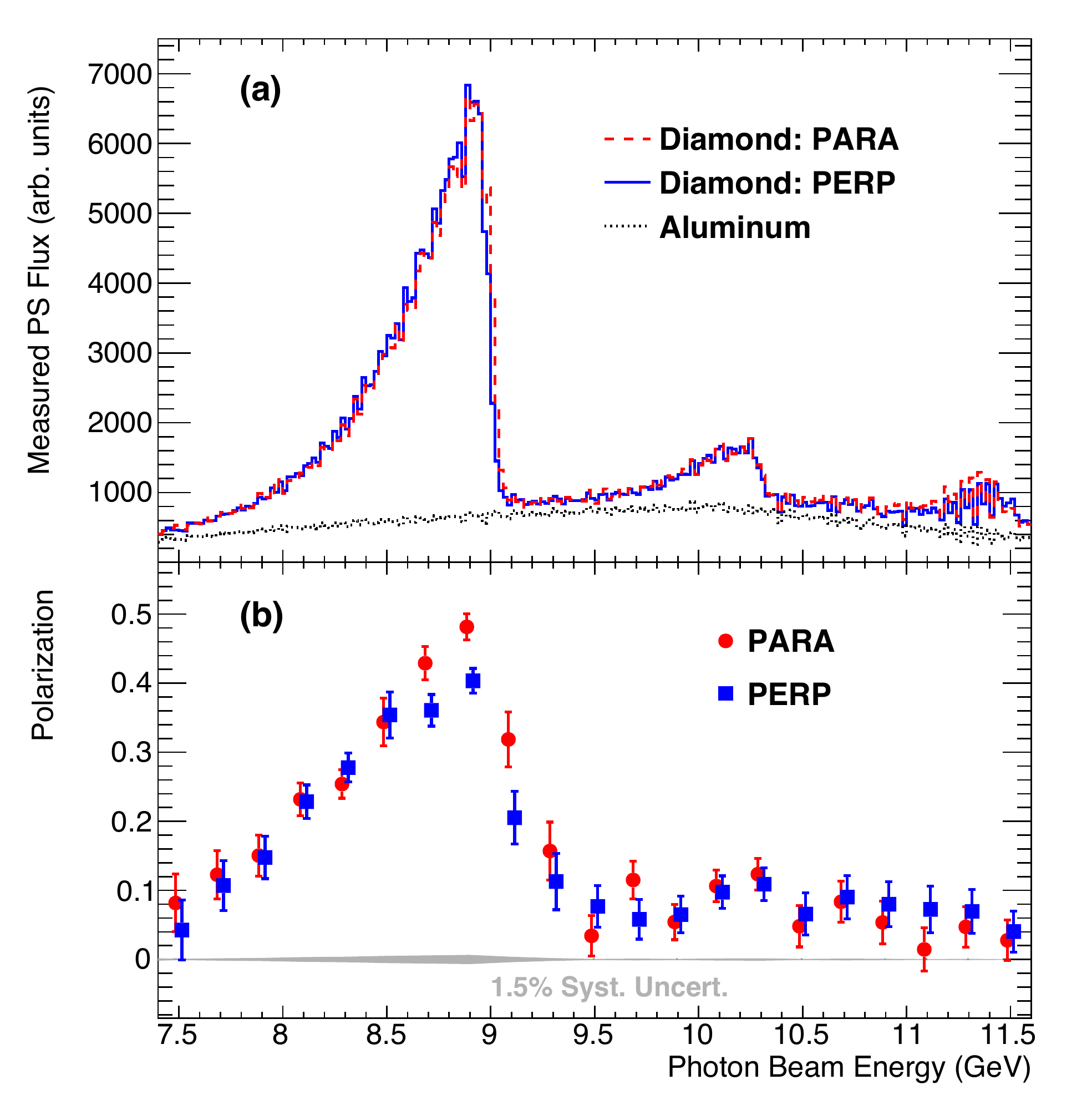}
\end{center}

\caption{(Left) Schematic of the GlueX detector highlighting the major components.  The DIRC is expected to be installed by 2019.  (Right) Beam properties from Ref.~\cite{pi0sigma}: Measured beam flux without acceptance corrections (top), average beam linear polarization (bottom).}
\label{fig:detector}
\end{figure}

The GlueX detector is schematically illustrated in Fig.~\ref{fig:detector}~(left).  The GlueX detector components and performance have been previously described in detail~\cite{hadron2015}.  The following is an overview of the experimental setup. Electrons with a maximum energy of 12 GeV are extracted from CEBAF, and pass through a radiator where photons are produced via bremsstrahlung radiation.  The electrons then pass into a dipole magnet, which allows for the tagging of the bremsstrahlung photon energies by the detection of the scatted electrons in a pair of hodoscopes.  For 12 GeV electrons, the Tagger Microscope (TAGM) covers the 8.2--9.2 GeV range in 10~MeV increments, while the rest of the range down to 7 GeV is covered by the Tagger Hodoscope (TAGH) in $10-25$~MeV increments.  For some of the earlier running at lower rates, an additional section of the TAGH with 50\% coverage down to 3~GeV was also used.

The primary type of bremsstrahlung radiator used in GlueX is thin diamond crystals, which are aligned to produce coherent bremsstrahlung radiation in the range covered by the TAGM, peaking at $\sim9$~GeV. 
The diamond can be rotated to change the direction of the linear polarization, and data are taken with orthogonal polarization directions.  The polarization directions are conventionally chosen to be parallel and perpendicular to the floor of the detector hall, but half of the 2017 data was taken with these polarization angles rotated by $45^\circ$ for systematic studies.

The beam of bremsstrahlung photons proceeds down an 80~m beampipe, after which it is collimated and the beam properties are monitored using well-known QED processes.  The beam is incident on 75~$\mu$m thick Beryllium radiator, after which electron-position pair conversions $\gamma + \mathrm{Be} \to e^+e^- + (\mathrm{Be})$ are measured in the two-armed Pair Spectrometer (PS)~\cite{gluexps}, to determine the flux and energy spectrum of the beam.  An example of the electron-positron pair spectrum measured in the PS before acceptance corrections from 2016 data is shown in the top panel of Fig.~\ref{fig:detector}~(right). The peaks due to coherent bremsstrahlung radiation in the diamond radiator are clearly seen, with the main peak near 9~GeV, as expected. In this panel, for comparison, the measured spectrum from a thin aluminum radiator are shown, in which only incoherent bremsstrahlung radiation is produced.  The polarization of these photons is determined by measuring the triplet production process $\gamma + e^- \to e^+e^- + e^-$, where the electron-positron pair is measured in the PS, and third, recoiling atomic electron is measured in the Triplet Polarimeter (TPOL), which consists of an azimuthally-segment silicon detector.  The lower panel of Fig.~\ref{fig:detector}~(bottom right) shows the measured polarization of the photon spectrum in the top panel, which averages $\sim40\%$ in the primary coherent peak, consistent with expectations.

The photon beam continues downstream and is incident on a liquid hydrogen target 30~cm in length, which sits in the middle of the main GlueX detector.  The detector features a solenoidal magnet providing a magnetic field of about 2~T in the direction of the beam, and nearly uniform detection of charged and neutral particles in the angular range $\theta = 1^\circ - 120^\circ$.  Charged particle detection is provided by the Central Drift Chamber (CDC)~\cite{gluexcdc}, which uses straw tubes to provide position measurements with 150~$\mu$m resolution in the $r-\phi$ plane and on the mm scale in the $z$-direction, and the Forward Drift Chamber (FDC)~\cite{gluexfdc}, which uses planar drift chambers with cathode and anode readouts to deliver a 3-dimensional position with 200~$\mu$m resolution.  Neutral particle detection is provided by the Barrel Calorimeter (BCAL)~\cite{gluexbcal}, which consists of a lead-scintillating fiber matrix covering polar angles from $12^\circ$ to $160^\circ$, and the Forward Calorimeter (FCAL)~\cite{gluexfcal}, which consists of a wall of 2800 lead-glass crystals covering polar angles from $1^\circ$ to $12^\circ$.  The expected energy resolutions of the BCAL and FCAL are $\sigma_E / E \approx 5.4\% \oplus 2.3\%$ and  $\sigma_E / E \approx 5.6\% \oplus 3.5\%$, respectively.  Particle identification is provided by time-of-flight measurements between the scintillator-based Start Counter (SC) surrounding the target, which has a time resolution of $\sim250$~ps, and either the BCAL, with a time resolution of $\sim150$~ps, or the Time-of-Flight wall (TOF), which consists of two layers of scintillator paddles located right in front of the FCAL, which has a time resolution of $100$~ps.  Ionization measurements in the CDC can also be used to distinguish charged pions and protons with $p<1$~GeV.  All detectors have been commissioned and are operating at or near design performance.

\begin{figure}[!tb]
\begin{center}
\includegraphics[width=1.9in]{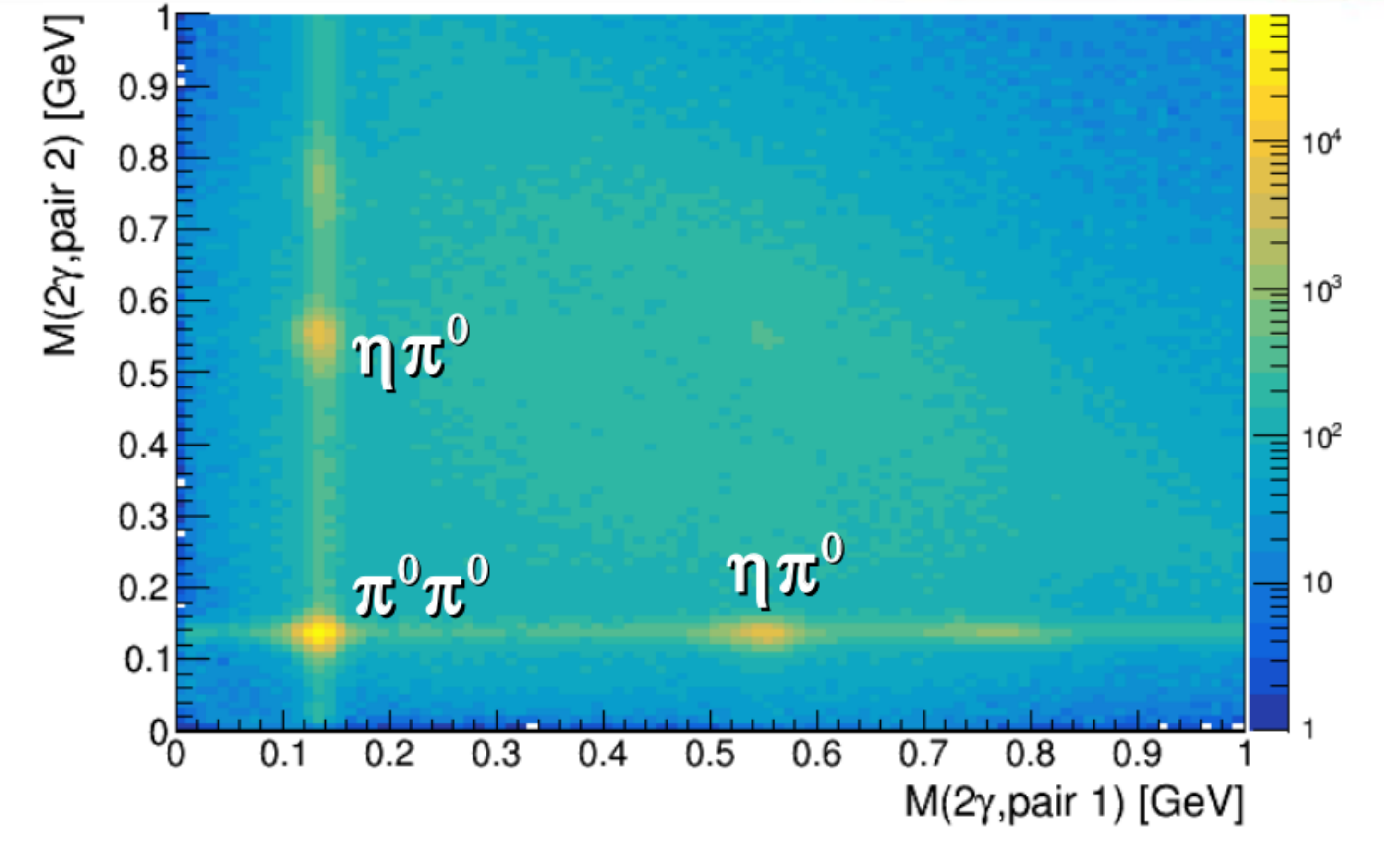}
\hspace{5pt}
\includegraphics[width=3.8in]{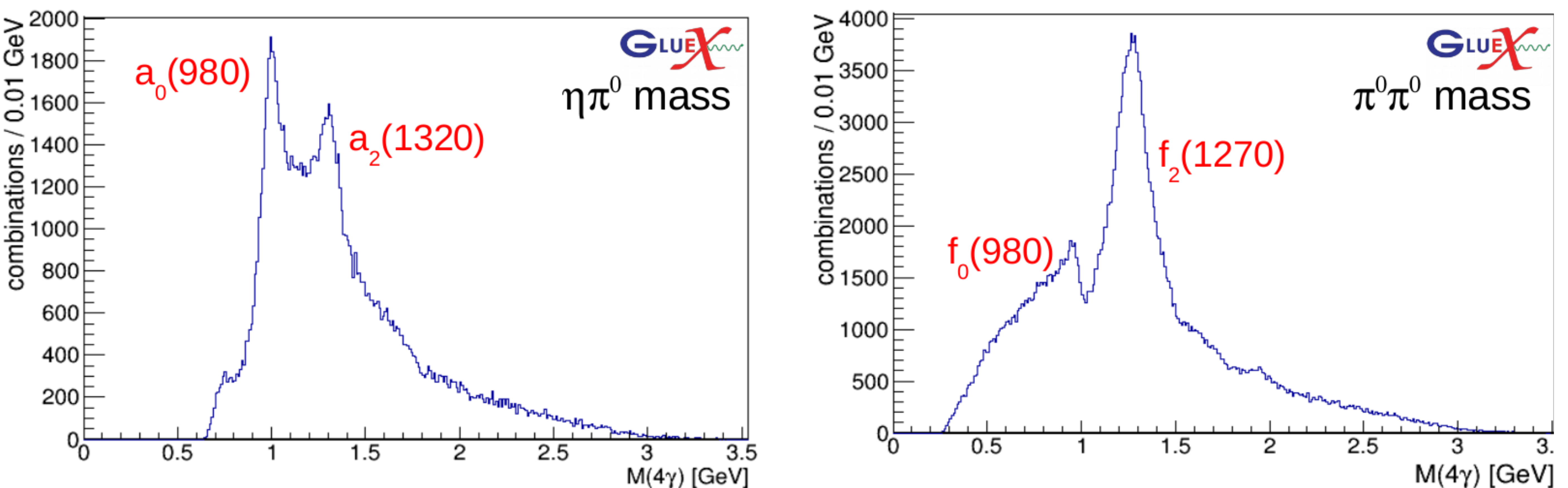}
\end{center}

\caption{Mass distributions from $\gamma p \to p + 4\gamma$ events. Locations of several well-known mesons are indicated in the center and right-hand panels.}
\label{fig:4gamma}
\end{figure}

\begin{figure}[!tb]
\begin{center}
\includegraphics[width=1.8in]{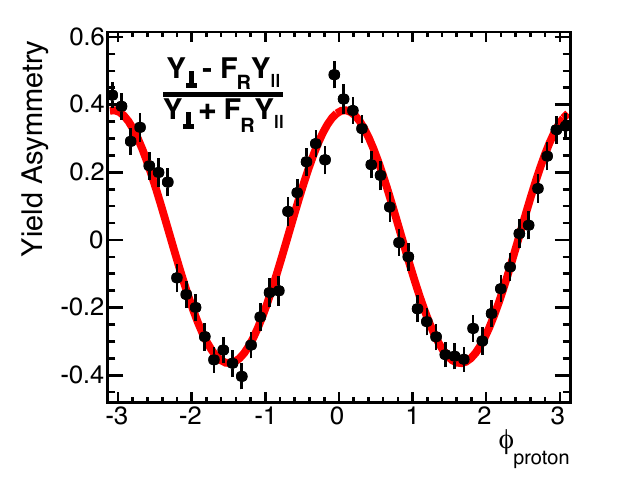}
\raisebox{5pt}{\includegraphics[width=1.95in]{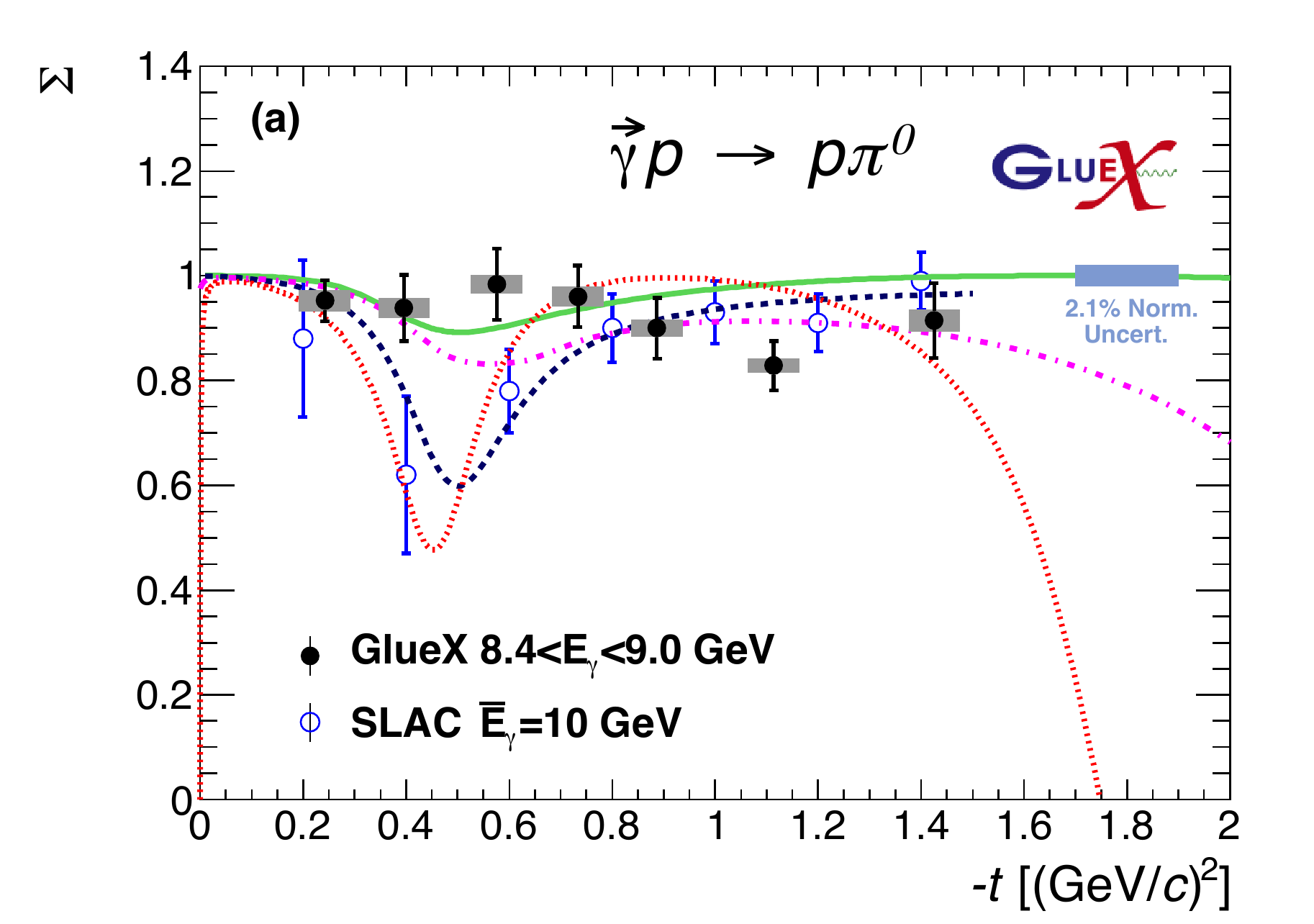}}
\includegraphics[width=2.in]{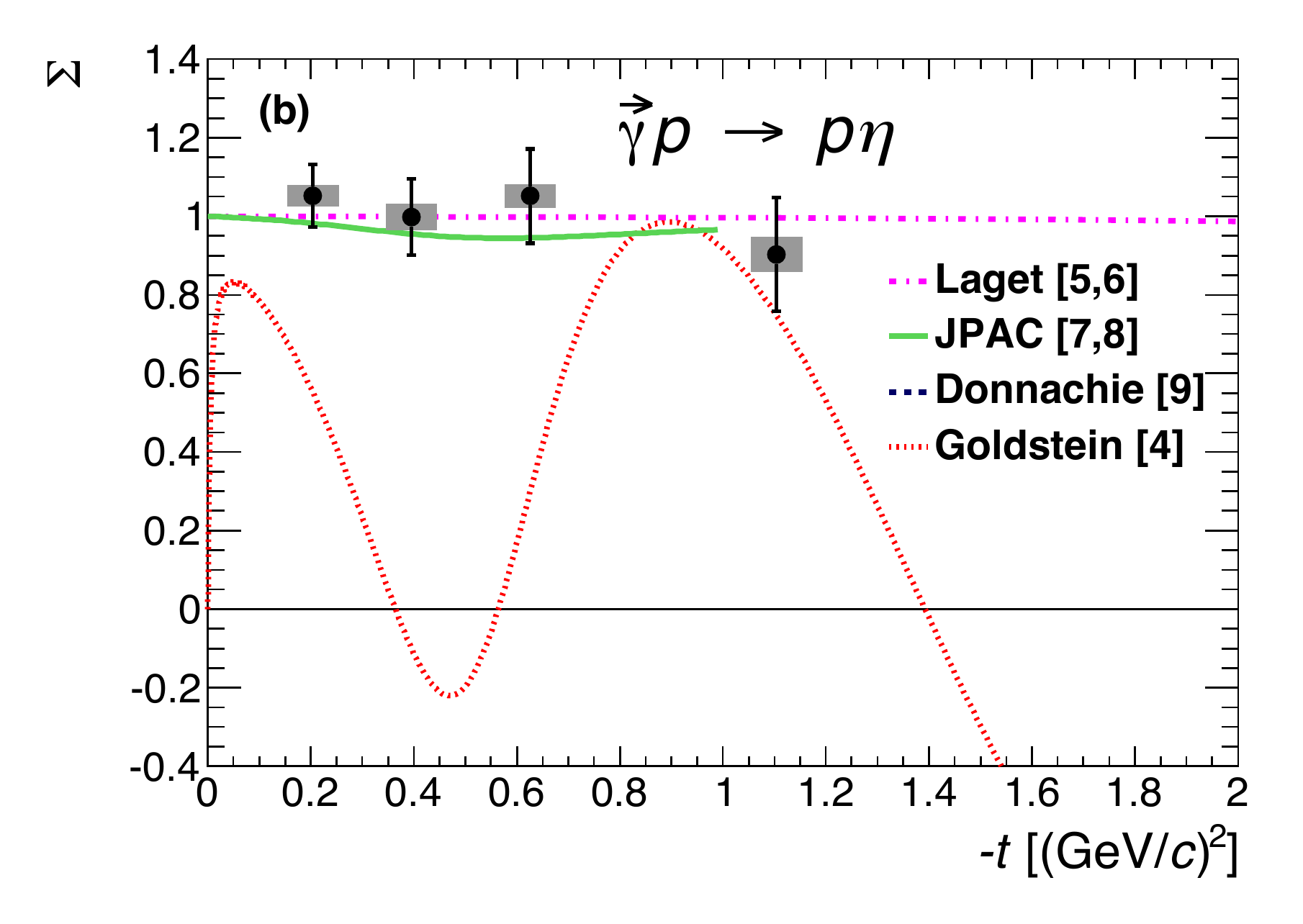}
\end{center}

\caption{From Ref.~\cite{pi0sigma}: (Left) Total normalized yield asymmetry for $\gamma p \to p \pi^0$ from 2016 GlueX data.  The red curve illustrates the fit used to extract the beam asymmetry $\Sigma$. The $\Sigma$ is shown as a function of Mandelstam$-t$ for the production of $\pi^0$ (middle) and $\eta$ (right), along with several model predictions~\cite{jpacpi0,jpaceta,laget,goldstein,donnachie}.}
\label{fig:pi0}
\end{figure}

\section{Early Spectroscopy and Beam Asymmetry Measurements}

Hybrid mesons are generally expected to produced in the same final states as other, normal mesons, and their identification and determination of their $J^{PC}$ will require amplitude analyses of these final states.  There is little existing photoproduction data at $E_\gamma \approx 9$~GeV, our current knowledge coming mostly from photoproduction experiments at SLAC in the 1970's and early 1980's.  Therefore, our first steps are to study the mechanisms of meson photoproduction in detail and to identify known mesons in order to build the theoretical models and understanding of the GlueX detector needed to move on to the search for hybrid mesons.

Final states containing neutral particles are particularly interesting, as previous experiments had low efficiency for these particles and they were studied with limited precision.  To determine the capability of GlueX to reconstruct these final states, several all-neutral final states have been studied.  The reaction $\gamma + p \to p + 4\gamma$ is shown as an example. The proton is reconstructed as a charged track in the CDC with ionization loss consistent with that expected by a proton, and the photons as calorimeter showers of energy $>100$~MeV not associated with a charged particle track.  A kinematic fit to the $\gamma + p \to p + 4\gamma$ hypothesis was done for all identified combinations of proton and shower candidates in an event, and combinations with a fit probability of $>1\%$ were kept.  For the events that passed these selections, Fig.~\ref{fig:4gamma}~(left) shows a 2-dimensional plot of the invariant masses for all possible 2-photon combinations. Clear signals for $\pi^0\pi^0$ and $\eta\pi^0$ production are seen, where the $\pi^0$ and $\eta$ are identified in the decay into two photons. Fig.~\ref{fig:4gamma}~(center, right) shows the distribution of $M(4\gamma)$ when the $\pi^0\pi^0$ and $\eta\pi^0$ decays are selected.  In these distributions, features at the locations of several well-established states are seen, including peaks at the positions of $f_2(1270)$, $a_0(980)$, and $a_2(1320)$, and a dip in the $\pi^0\pi^0$ spectrum typically corresponding to the contribution of the $f_0(980)$.  These spectra show many features with a large number of events and therefore the potential for a rich spectroscopy, and are under further study on the path towards a full amplitude analysis.  

Another interesting final state is $\gamma + p \to p + 5\gamma$, in which a sample of over 10,000 $\gamma p \to p + b_1^0(1235)$, $b_1^0(1235) \to \omega \pi^0$, $\omega \to \pi^0 \gamma$ were reconstructed and are being studied.  This reaction is interesting not just because of the identification of another known meson, but also as a step towards studying the $b_1(1235)\pi$ final state, which is one of the promising final states for the identification of hybrid mesons easily accessible at GlueX.

Another piece of the puzzle is understanding the process of meson photoproduction.  The first steps being made towards this understanding are comprehensive studies of single meson photoproduction.  Insight into the production of single pseudoscalar mesons, $\gamma p \to p + (\pi^0,\eta,\eta',...)$, can be obtained by measuring the beam asymmetry $\Sigma$ as a function of the proton's momentum transfer (i.e., the Mandelstam $t$ variable).  The beam asymmetry is related to the azimuthal dependence of the production cross section, $\sigma \sim (1 - P_\gamma \Sigma \cos(2\phi))$, where $P_\gamma$ is the percentage of photon linear polarization, and $\phi$ is the difference between the azimuthal angle of the scattered proton and the photon linear polarization angle.  The $\Sigma$ parameter can then be extracted from a fit to the normalized azimuthal yield asymmetry, as illustrated in Fig.~\ref{fig:pi0}~(left).  Using the 2016 data set, the beam asymmetry $\Sigma$ is measured for the reactions $\gamma p \to p + (\pi^0,\eta)$, and the published results are shown in Fig.~\ref{fig:pi0}~(center,right)~\cite{pi0sigma}.  These measurements are much more precise than the only existing measurement for $\gamma p \to p \pi^0$ and are the first for  $\gamma p \to p \eta$ at these energies.  These measurements are also compared to various models, and in the context of the JPAC models~\cite{jpacpi0,jpaceta}, the fact that $\Sigma \sim 1$ indicates the dominance of vector particle exchange.

These single pseudoscalar production measurements are being extended to multiple decays of the $\eta$ and $\eta'$ mesons with the 2017 data.  Also, studies of spin-density matrix elements in the reactions $\gamma p \to p + (\rho,\omega,\phi)$ are in progress, which promises to further extend our understanding of the production amplitudes in these reactions.

%

\begin{figure}[!tb]
\begin{center}
\rotatebox{90}{\includegraphics[width=1.6in]{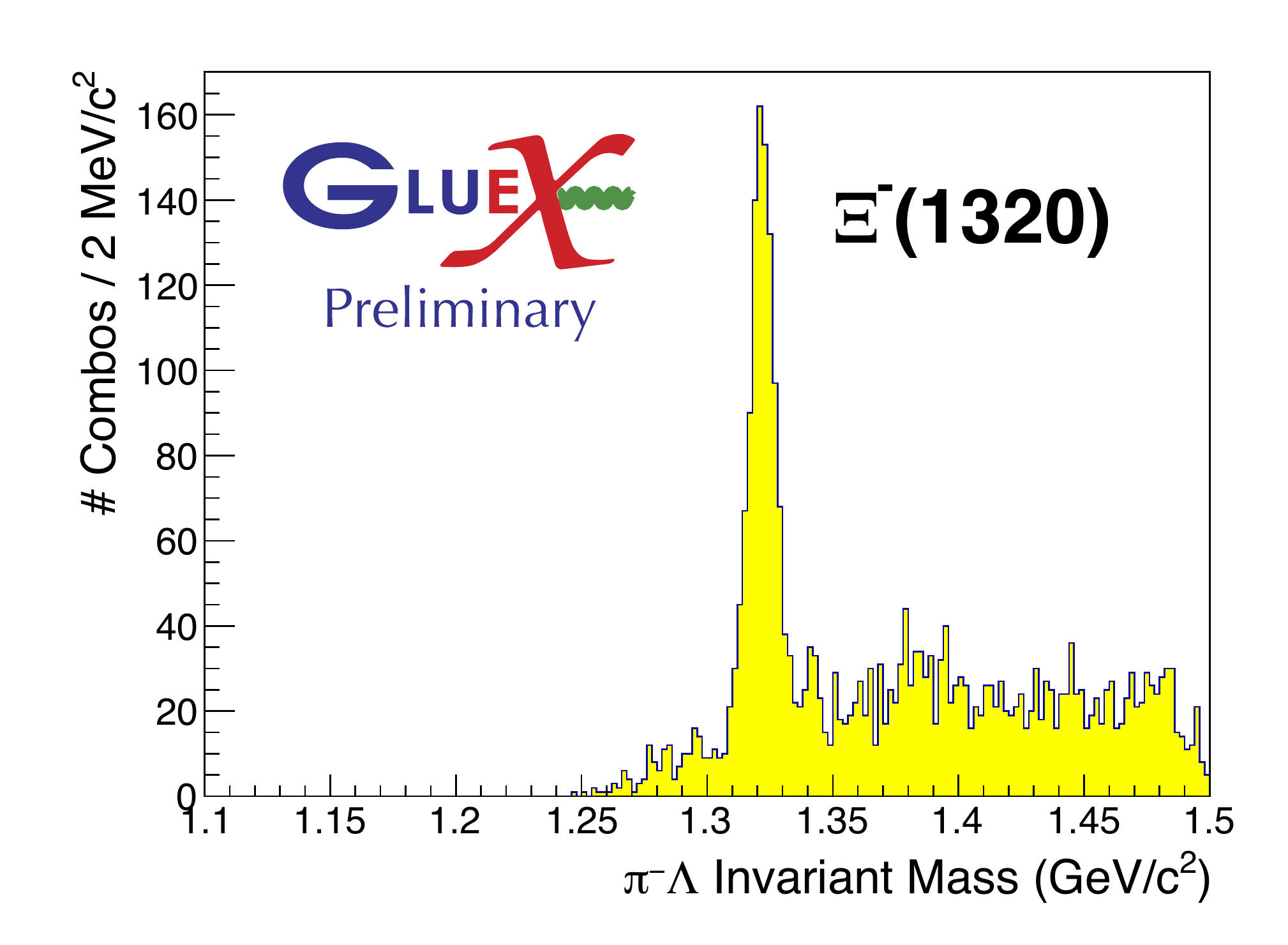}}
\rotatebox{90}{\includegraphics[width=1.6in]{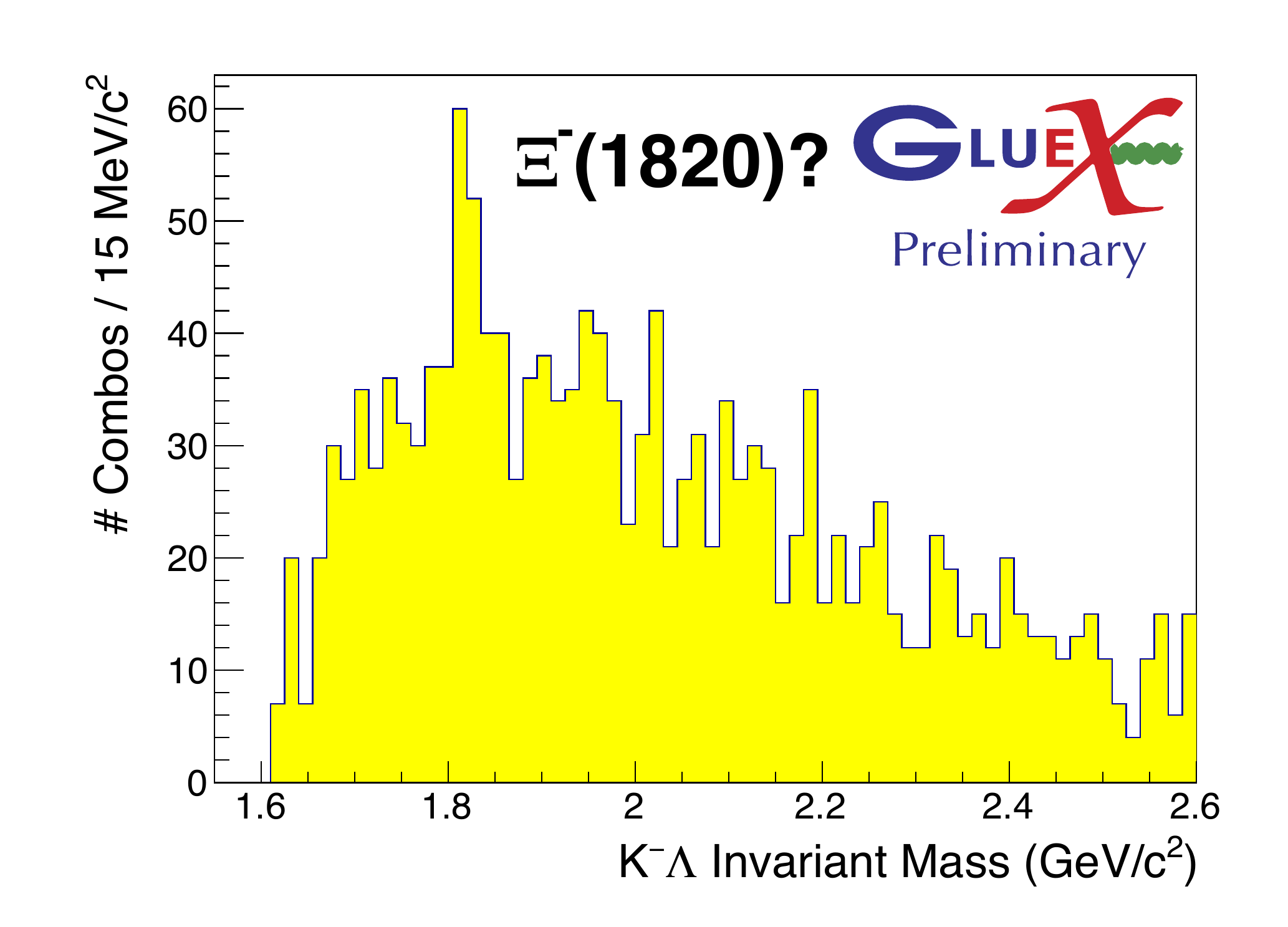}}
\end{center}

\caption{Preliminary mass spectra from fully-reconstructed events as described in the text, illustrating photoproduction of Cascade baryons: (left) $\pi^-\Lambda$ invariant mass; (right) $K^-\Lambda$ invariant mass.}
\label{fig:cascade}
\end{figure}

\section{Search for Photoproduction of Cascade Baryons}

Besides the main program of exotic meson spectroscopy, the large data sets collected by GlueX present an opportunity to study many other aspects of hadronic physics.  One example is the spectroscopy of the doubly-strange baryons, $\Xi^0$ ($uss$) and $\Xi^-$ ($dss$).  The $\Xi$ baryons are experimentally interesting since they are expected to have many narrow states~\cite{cascadelattice}, which should yield a unique probe of the strong interaction.  However, only 6 $\Xi$ states are considered well-established (3- or 4-stars in the PDG), there is almost no experimental data on their have their spin-parities, and no new information on their properties has been established in nearly 30 years~\cite{pdg}.  An earlier search for $\Xi$ photoproduction at CLAS with the 6~GeV electron beam from CEBAF observed production of the two lowest mass states, $\Xi(1320)$ and $\Xi(1530)$, but no evidence for any excited states.  With the larger phase space available to us at GlueX with the upgraded 12~GeV beam, the first step to search for the photoproduction of these states has been taken.

It is reasonable to expect that $\Xi$ states to be produced in photoproduction through the decay of excited hyperon states.  
Using GlueX data taken in 2017, the production of the $\Xi$ ground state is searched for in the reaction $\gamma p \to K^+K^+ \Xi^-$, $\Xi^- \to \Lambda \pi^-$, $\Lambda \to p \pi^-$.  This reaction is fully reconstructed using stringent time-of-flight cuts for particle identification, and a four-momentum conserving kinematic fit, in which the vertices for each step of the reaction are individually constrained, to take into account the non-negligible flight distances of the hyperons.
The $\Lambda\pi^-$ invariant mass spectrum is shown in Fig.~\ref{fig:cascade}~(left), with a clear signal for the production of the $\Xi(1320)$.  

The production of excited $\Xi^*$ is then searched for in the similar reaction, but with the decay $\Xi^{*-} \to \Lambda K^-$, reconstructed in a similar fashion as the previous reaction, and including a veto for events with a $K^+K^-$ mass consistent with that of the $\phi$ meson.
The $\Lambda\pi^-$ invariant mass spectrum is shown in Fig.~\ref{fig:cascade}~(right), with a peak at a mass of $\sim1.82$~GeV.  The mass of this peak is consistent with that of the $\Xi(1820)$ and would correspond to the first evidence for photoproduction of excited $\Xi$ states.  The full GlueX data for these reactions is currently being analyzed in detail, but these preliminary measurements seem to open the door to a broader program of $\Xi$ baryon spectroscopy at GlueX.

\begin{figure}[!tb]
\begin{center}
\includegraphics[width=2.3in]{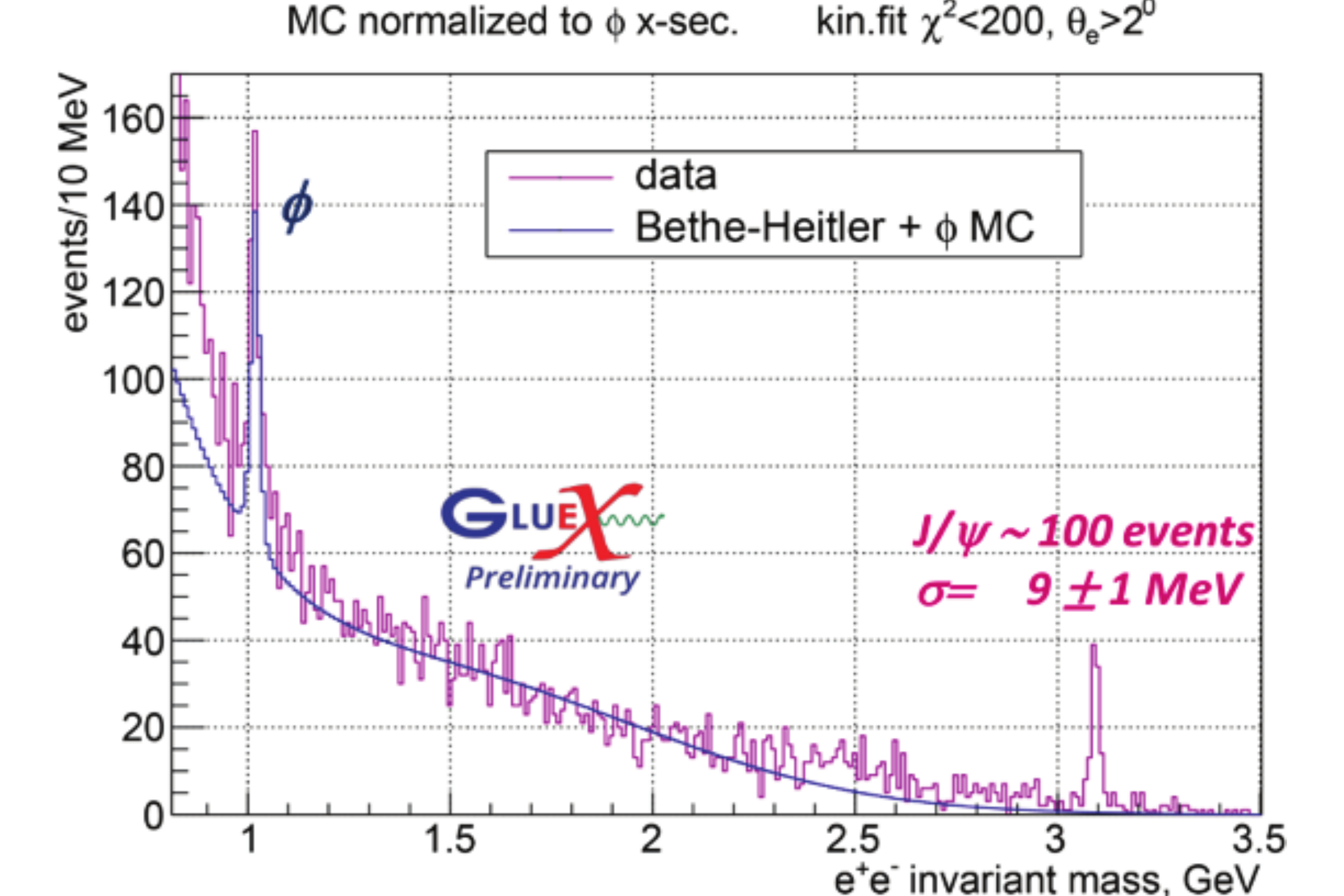}
\includegraphics[width=1.3in]{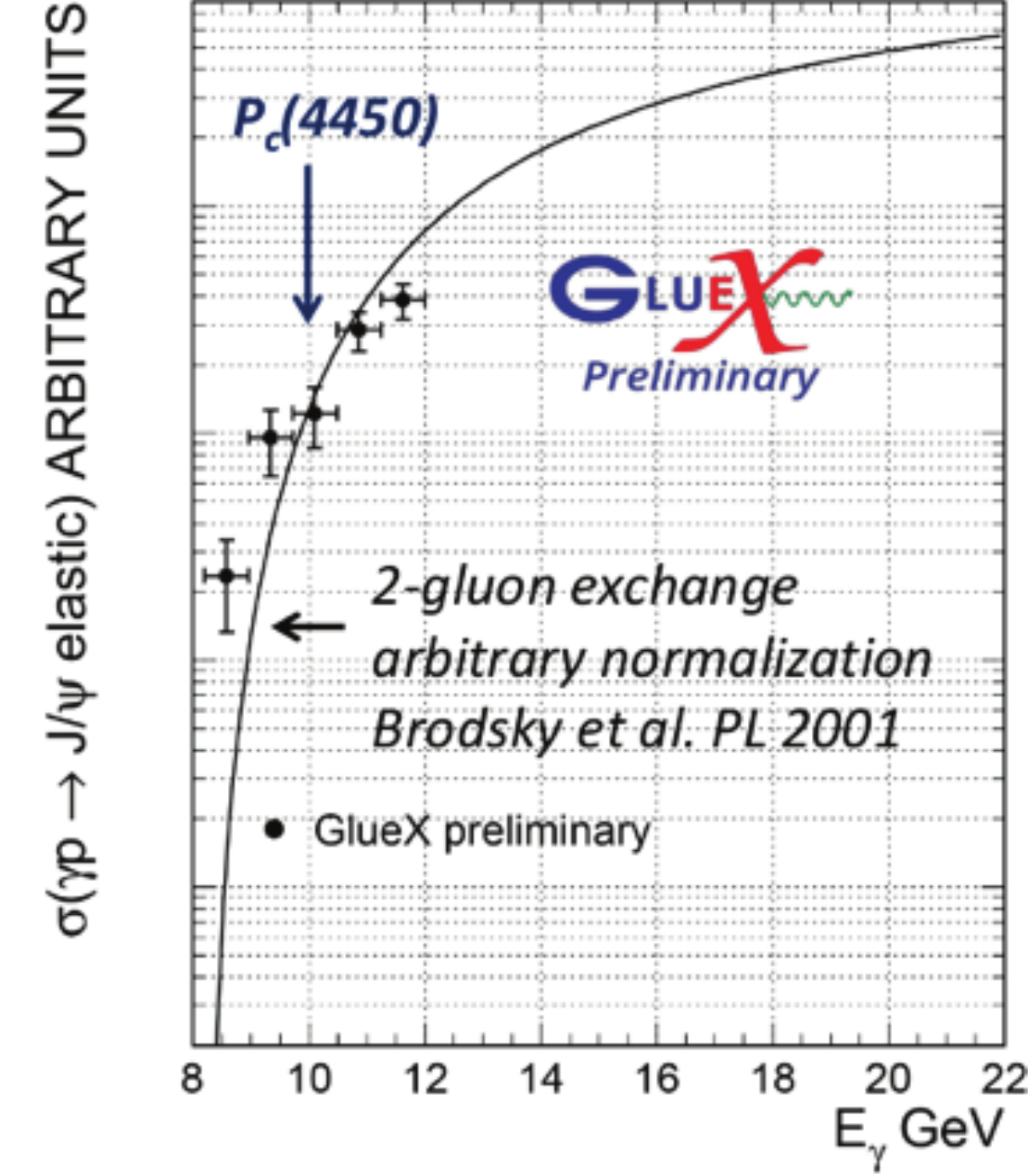}
\includegraphics[width=2.2in]{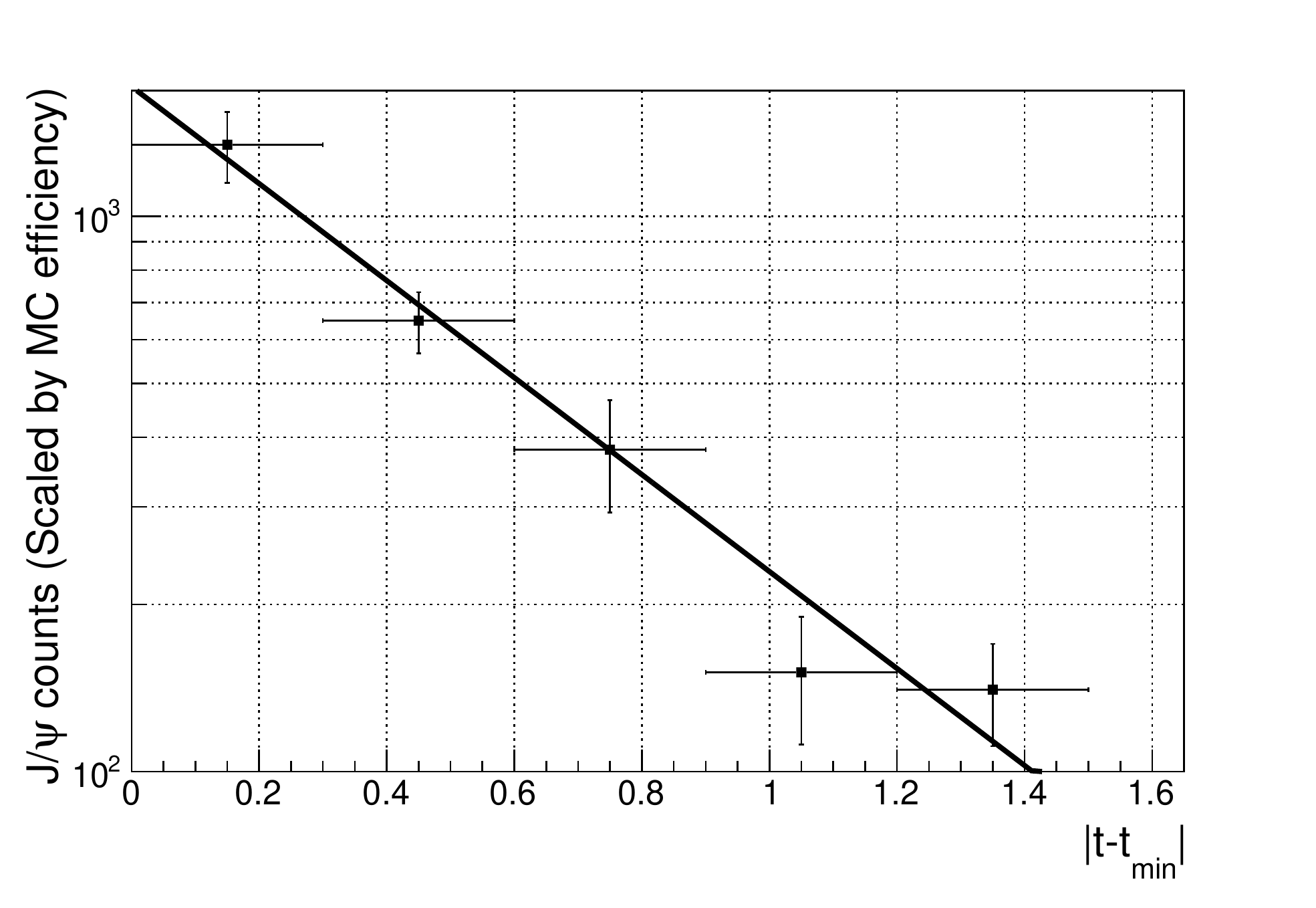}
\end{center}

\caption{Preliminary results for $J/\psi$ production: (Left) Invariant mass spectra of $e^+e^-$ showing peaks corresponding to $J/\psi$ and $\phi$ production.  (Middle) Arbitrarily normalized variation of $J/\psi$ yield as a function of beam energy, illustrating the statistical precision of the 2016 data. (Right) Efficiency corrected t-distribution of $J/\psi$ events along with an exponential fit curve.}
\label{fig:jpsi}
\end{figure}

\section{Charmonium Photoproduction Near Threshold}

The study of heavy quarkonium production has a long history of contributing to the development of QCD.  Photoproduction of $J/\psi$ near threshold provides an unusually clean laboratory to study charmonium production due to the electromagnetic photon probe and the lack of feed-down contributions from the decays of charmonium states.  The study of $J/\psi$ photoproduction can provide information on the gluonic structure of the nucleus~\cite{brodsky,kharzeev}, for example, through measuring the contribution of different production mechanisms (e.g. leading-order and higher-twist contributions)~\cite{brodsky}.  This process can also allow us to study the photocouplings of the $P_c^+$ states observed in LHCb~\cite{wang,kubarovsky,karliner,jpacjpsi}.  Little data exists near threshold, with the only experiment at the $E_\gamma < 12$~GeV range accessible at GlueX being a 1975 measurement performed at Cornell~\cite{cornelljpsi}. The initial GlueX running will allow first measurements of $d\sigma/dE_\gamma$ and the t-slope of $J/\psi$ production.

The reaction $\gamma p \to J/\psi p$, $J/\psi \to e^+e^-$ is used, where the electrons are identified by their electromagnetic showers in the calorimeters, and a kinematic fit is performed to the full reaction.  Fig.~\ref{fig:jpsi}~(left) shows the dielectron mass spectrum from the 2016 data, which is well described by simulation down to $M\sim1$~GeV.  Approximately 100 $J/\psi$ are observed at a mass consistent with the well-known $J/\psi$ mass.  Fig.~\ref{fig:jpsi}~(center) shows preliminary results for the arbitrarily normalized event rate as a function of the incident beam energy to illustrate the statistical precision of the data.  Initial GlueX running is expected to provide up to an order of magnitude more events than the 2016 data, with the full GlueX data yielding an increase of two orders of magnitude.

The value of the $t-$slope of the reaction is necessary for measuring these total cross sections, and a new measurement is interesting since the previous measurements at the lowest energies of $2.9\pm0.3$~GeV$^{-2}$ at SLAC with $E_\gamma = 19$~GeV~\cite{slacjpsi}, and of $1.13\pm0.18$~GeV$^{-2}$ at Cornell with $E_\gamma\sim11$~GeV~\cite{cornelljpsi} disagree with each other.  Fig.~\ref{fig:jpsi}~(right) shows preliminary results for the $t-$slope using GlueX data from 2016 and 2017, which leads to a slope of $2.01\pm0.36(\mathrm{stat})$~GeV$^{-2}$.

\section{Summary}

The GlueX detector has been successfully commissioned and has started its physics data taking campaign in Spring 2017.  Along with data taken in a commissioning run in 2016, these data are being analyzed for a variety of physics topics.  Photoproduction mechanisms for single pseudoscalar meson and vector meson production are being studied as a first step towards amplitude analysis of the light meson spectrum. The first observations of Cascade baryon and $J/\psi$ photoproduction at GlueX have been made, which show good prospects for a broad physics program with initial GlueX data.  After the completion of initial GlueX running, expected in 2019, several upgrades are planned including the addition of a DIRC detector in the forward region to improve $\pi/K$ separation and enhance the program of strange hadron spectroscopy, and support for a factor 5 or more higher luminosity.

This material is based upon work supported by the U.S. Department of Energy, Office of Science, Office of Nuclear Physics under contract DE-AC05-06OR23177.

\end{document}